\documentclass[12pt,a4paper]{article}
\usepackage{dfttrob}
\usepackage{times}
\usepackage{amsmath}
\usepackage{latexuseful2e}
\usepackage{graphicx}

\newcommand{\citelow}[1]{\cite{#1}}
\newcommand{\abstracts}[1]{\begin{center}\textbf{Abstract}\end{center}
\begin{quote}#1\end{quote}}

\newcommand{\pfrac}[3]{\left(\frac{\partial #1}{\partial #2}\right)_{\!\!#3}}

\begin{document}
\dfttnum{DFTT 30/2002}

\title{Power series distributions in clan structure analysis: new
observables in strong interactions}

\author{R. UGOCCIONI AND A. GIOVANNINI\\
\small\itshape
Dipartimento di Fisica Teorica and I.N.F.N. - sezione di Torino \\
\small\itshape via P. Giuria 1, I-10125 Torino, Italy}

\maketitle

\abstracts{
We present a new thermodynamical approach to multiparticle production
in high energy hadronic interactions, making use of the formalism of
infinitely divisible power series distributions. This approach allows
us to define new observables, linked to the system fugacity, which
characterise different classes of events. 
}

\vspace{2cm}
\begin{center}
Talk presented by R. Ugoccioni at the
X International Workshop on Multiparticle Production
``Correlations and Fluctuations 2002'', 
Crete, Greece, 8-15 June 2002.
\end{center}
\newpage

\section{Introduction}

The phenomenological analysis of many-particle final states in
hadron-hadron collisions in the GeV region has been successfully
carried out\cite{combo:prd} using a two-component model: each event is
assigned to one of two classes, called `soft' and `semi-hard', which
correspond to events without mini-jets and events with mini-jets,
respectively.
We also assume that the multiplicity distribution (MD) in each class
is described by a negative binomial (Pascal) distribution (NBD), of
course with different parameters $\nbar$, $k$ in each class.  This
model was successful in describing the shoulder in MD's, the
oscillations of high rank moments thereof\cite{combo:prd} and also
forward backward multiplicity correlations.\cite{RU:FB} There are also
experimental indications that the two classes behave differently in
the TeV region.\cite{CDF:soft-hard}

In order to extrapolate to the LHC region we have to make some
assumptions on the behaviour of the parameters of the two NBD's
which we summarise as follows:

\begin{itemize}
\item The overall average multiplicity grows as $\ln^2\sqrt{s}$.

\item The soft component average multiplicity grows as $\ln\sqrt{s}$ and
the MD obeys KNO scaling (thus $k_{\text{soft}}$ is
approximately constant.)

\item The average multiplicity in the semi-hard component is
approximately twice as large as in the soft one.

\item Three scenarios have been examined for the behaviour of
$k_{\text{semi-hard}}$:
  \begin{enumerate}
	  \item same behaviour as for the soft component, i.e., 
it is constant (therefore KNO scaling is satisfied);
  	\item $k_{\text{semi-hard}}^{-1} \approx \ln\sqrt{s}$, implying a
strong violation of KNO scaling;
  	\item it follows a QCD-inspired behaviour; KNO scaling is attained
only asymptotically. This last scenario is intermediate between the 
first two. 
	\end{enumerate}
\end{itemize}

Of course, using NBD's means we can explore the clan 
structure:\cite{AGLVH:1,AGLVH:4} the average number of clans $\Nbar$ and the
average number of particles per clan are defined by
\begin{equation}
	\Nbar = k \ln \left( 1 + \frac{\nbar}{k} \right) ;
	\qquad
		\nc = \frac{ \nbar }{ k \ln \left( 1 + \nbar/k \right) } .
\end{equation}

It turns out that the second and third scenarios show a number of
clans which is rapidly decreasing with c.m.\ energy (accompanied by a
fast increase of the average number of particles per clan). This is
surprising, and in this paper we try to understand 
the implications of this result 
at parton level using thermodynamical concepts.

To connect the hadronic and partonic levels we use the generalised
local parton-hadron duality (GLPHD),\cite{AGLVH:2} which says that
all inclusive distributions are proportional at the two levels of
investigation:
\begin{equation}
	Q_{n,\text{hadrons}}(y_1,\dots,y_n) =
		\rho^n Q_{n,\text{partons}}(y_1,\dots,y_n) ,
\end{equation}
which corresponds for NBMD parameters to
\begin{equation}
 k_{\text{hadron}} = k_{\text{parton}} ,\qquad
	\nbar_{\text{hadron}} = \rho\; \nbar_{\text{parton}} .
\end{equation}
GLPHD will be applied separately to soft and semi-hard components.

\section{A new thermodynamical approach}

The thermodynamical approach to multiparticle production has a long
history which cannot be summarised here. We would just like to
attract the reader's attention to the result\cite{ScalapinoSugar} that, to
leading order in the allowed rapidity range, the 
generating function (GF) for the MD has the
form of an infinitely divisible distribution (IDD).

Keeping in mind this result, we propose the following approach.

The partition function in the canonical ensemble, $Q_n(V,T)$, 
for a system with $n$ particles, volume $V$
and temperature $T$, is linked to
the partition function in the grand-canonical ensemble with fugacity
$z$, ${\cal Q}(z,V,T)$,
by the well known relation
\begin{equation}
	{\cal Q}(z,V,T) = \sum_n z^n Q_n(V,T) .
\end{equation}

Quite in general, in the grand-canonical treatment the probability
of finding $n$ particles in a system is given by
\begin{equation}
	p(n) = \frac{z^n Q_n(V,T)}{{\cal Q}(z,V,T)} . \label{eq:1}
\end{equation}
That is to say, for a thermodynamical system the MD belongs to the
class of power series distributions (PSD's), 
and is characterised indeed by the following form:
\begin{equation}
  p(n) = \frac{a_n b^n}{\gamma(b)} ,
\end{equation}
with constants $a_n$, $b$.

We therefore propose,
given a MD in power series form, the following correspondence
with Eq.~(\ref{eq:1}):
\begin{align}
        z        &\longleftrightarrow b ,\nonumber\\
        Q_n      &\longleftrightarrow a_n ,  \label{eq:identify}\\
        {\cal Q} &\longleftrightarrow \gamma(b) = p(0)^{-1} .\nonumber
\end{align}

When the PSD is also IDD, then we know it can be cast in the form of a
compound Poisson distribution, such that
\begin{equation}
  p(0) = e^{-\Nbar} .    \label{eq:IDD}
\end{equation}
In a two-step approach, $\Nbar$ is average number of objects
(clans) generated in the first step. This way of describing the
partonic cascade is well known:\cite{AGLVH:4}
the ancestors (first step) are
independent intermediate gluon sources;  
it is their thermodynamic
properties which we want to explore. 

In our thermodynamical approach, $\Nbar$ becomes of fundamental
importance since Eq.s~(\ref{eq:identify}) and (\ref{eq:IDD}) imply
\begin{equation}
	\Nbar = -\ln p(0) = \ln {\cal Q} ;
\end{equation}
all thermodynamical properties can be obtained by differentiating
$\Nbar$.

From the standard relation $P V = k_B T \ln {\cal Q}$, we obtain the
equation of state
\begin{equation}
	P V = \Nbar k_B T ,
\end{equation}
which says that clans form a classical ideal gas.

The negative binomial (Pascal) distribution
belongs to both classes, power series and IDD.
The standard form, from which the correspondence with the partition
function can be obtained, is the following:
\begin{equation}
	p(n) = \frac{k(k+1)\dots(k+n-1)}{n!} 
			\left( \frac{\nbar}{\nbar+k} \right)^n
			\left( \frac{k}{\nbar+k} \right)^k .
\end{equation}
The identification we propose in our approach is
\begin{align}
	a_n &= \frac{k(k+1)\dots(k+n-1)}{n!} ,\nonumber\\
	b   &=  \frac{\nbar}{\nbar+k} , \label{eq:bnk}\\
	\gamma(b) &= \left(\frac{k}{\nbar+k}\right)^{-k} .\nonumber
\end{align}

Notice that $k(V,T)$ is the canonical partition function for a system
with 1 particle; it is in our approach an unknown function of $V$ and
$T$.

Finally notice that $b$ is the fugacity $z$:
\begin{equation}
	z = \frac{\nbar}{\nbar+k} .
\end{equation}

When the ancestors are created early in the evolution, at larger
virtualities and with higher temperature, they tend to follow a
quasi-classical behaviour, as the production of a new ancestor is
competitive with the increase in gluon population within each clan.
This results in a relatively large value of the $k$ parameter,
\textit{i.e.}, a small amount of aggregation. 
When the number of partons per clan is very small (close to 1;
$k$ is very large) then
essentially each parton is a clan, and the equation of state reduces 
basically to that of an ideal gas (quasi-classical behaviour):
\begin{equation}
	P V \approx \nbar k_B T .
\end{equation}
Via GLPHD, we expect a similar situation to hold at hadron level. This
behaviour is qualitatively close to that of soft events as well as of
scenario-1 semi-hard events.

When the ancestors are created later in the evolution, at lower
virtualities and with lower temperature, they tend to remember their
quantum nature, as newly produced gluons prefer to stay together with
other clan members rather than initiate a new clan.
This results in a relatively small value of the $k$ parameter,
\textit{i.e.}, a larger aggregation and larger two-particle correlations.
When the number of partons per clan begins to grow, the equation of
state for partons becomes more and more different
(quasi-quantum behaviour), but that for clans remains that of an ideal
gas. 
\begin{equation}
	P V = \Nbar k_B T = k\ln\left( 1 + \nbar/k \right) k_B T.
\end{equation}
Via GLPHD, at hadron level we recognise the behaviour of scenario-2
and scenario-3 semi-hard events.


It is interesting now to calculate some thermodynamical quantities.

The Helmholtz free energy
can be rewritten in a form symmetric in $\nbar$ and $k$:
\begin{equation}
  -\frac{A}{k_B T} = \nbar\mu - P V
								 = \nbar \ln \left(1+\frac{k}{\nbar}\right) + 
											k \ln\left(1+\frac{\nbar}{k}\right)  .
\end{equation}

The average internal energy is
\begin{equation}
  \frac{U}{k_B T} = k_B T^2 \pfrac{\Nbar}{T}{b,V}
				= \Nbar \pfrac{\ln k}{\ln T}{V}  .
\end{equation}

The entropy is
\begin{equation}
	S =  \frac{ U-A }{ T }
    =  k_B\left\{ -\frac{A}{k_B T} + T \pfrac{k}{T}{V} 
								\ln\left( 1+\frac{\nbar}{k} \right)
				\right\} .
\end{equation}
which coincides with $-A/T$ in the limit of $(\partial k/\partial T)_V
\to 0$, since also $U\to 0$.

For further discussion of thermodynamical quantities, see 
Ref.~\citelow{RU:clanthermo}.

\section{Clan behaviour as a function of fugacity}
\begin{figure}
  \begin{center}
  \mbox{\includegraphics[height=0.92\textheight]{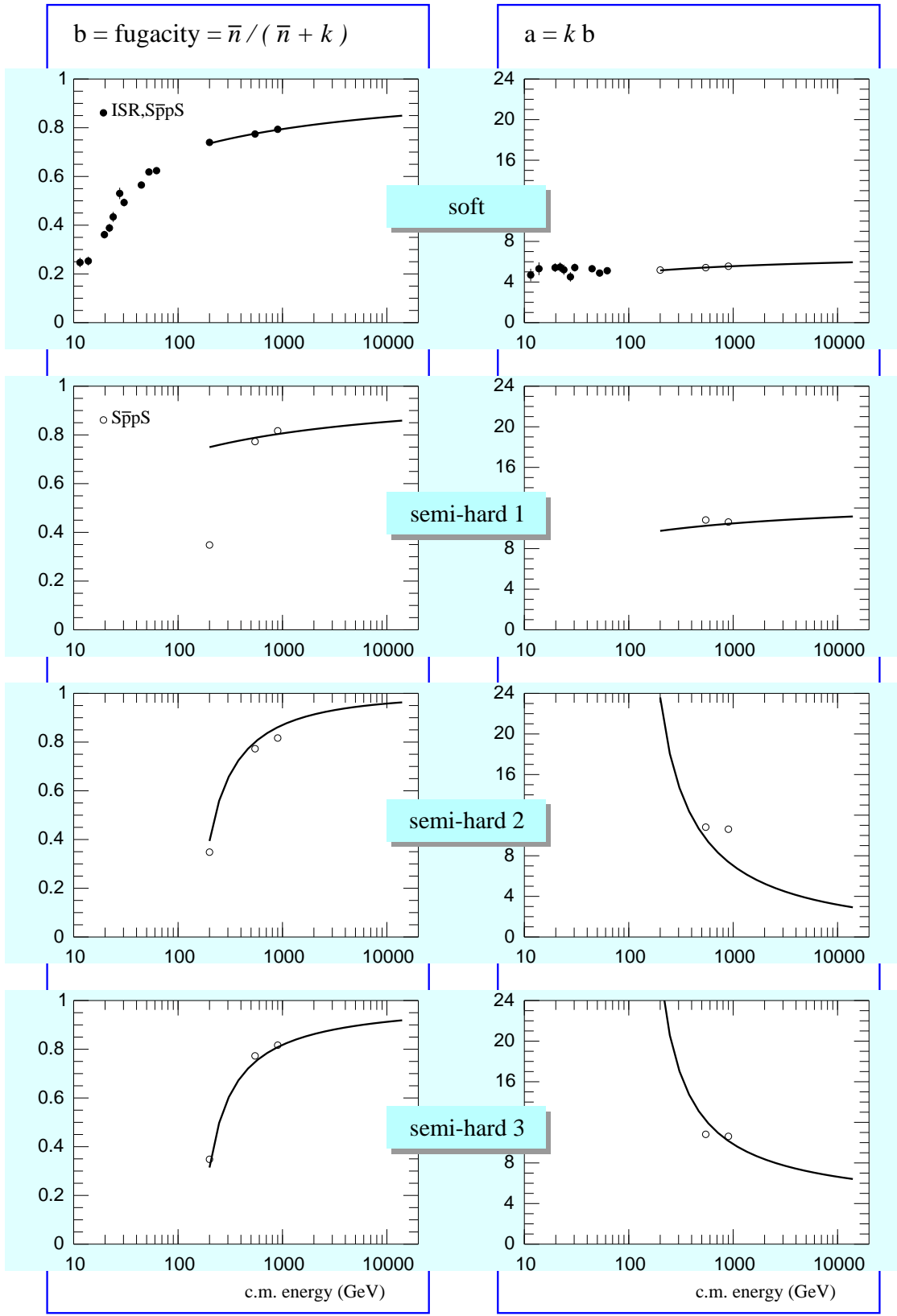}}
  \end{center}
  \caption{Fugacity and $a$ parameter 
	dependence on c.m.\ energy}\label{fig:fugacity}
  \end{figure}

Relying on GLPHD,
we analyse first experimental data on the fugacity and the related $a$
parameter: 
the NBD satisfies the recurrence relation
\begin{equation}
	\frac{ (n+1) p(n+1) }{ p(n) } = a + b n ,
\end{equation}
where
\begin{equation}
	a = \frac{ \nbar k }{ \nbar + k } ; \qquad b = \frac{ \nbar }{ \nbar
			+ k } .
\end{equation}
From Eq.~(\ref{eq:bnk}) it is seen that $b$ is the fugacity.  In
Figure~\ref{fig:fugacity} we show for each component and each scenario
the energy variation of the parameters $a$ and $b$.  
The points come from NB fits to experimental MD's, the lines show
the predictions from the extrapolation mentioned in the introduction.

The $a$ parameter
corresponds to the average multiplicity for a classical (Poisson)
system.  The relative behaviour of $b$ and $a=kb$ as the c.m.\
energy increases can be considered an indication of the relative
importance of a behaviour closer to a quantum one, i.e.\ harder, with
respect to a behaviour closer to a quasi-classical, i.e.\ softer, for
a class of events.  A very slow increase of $b$ with c.m.\ energy and
an almost constant behaviour of $a$ is the main characteristic
of the class of soft events and of scenario-1 semi-hard events.
A very fast decrease of $a$  in scenarios 2 and 3 and larger values of
the fugacity $b$ characterise harder events: the assumption of strong KNO
scaling violation for the semi-hard component (an extreme point of
view with respect to that of scenario 1) implies a completely new
panorama.

\begin{figure}
  \begin{center}
  \mbox{\includegraphics[height=0.92\textheight]{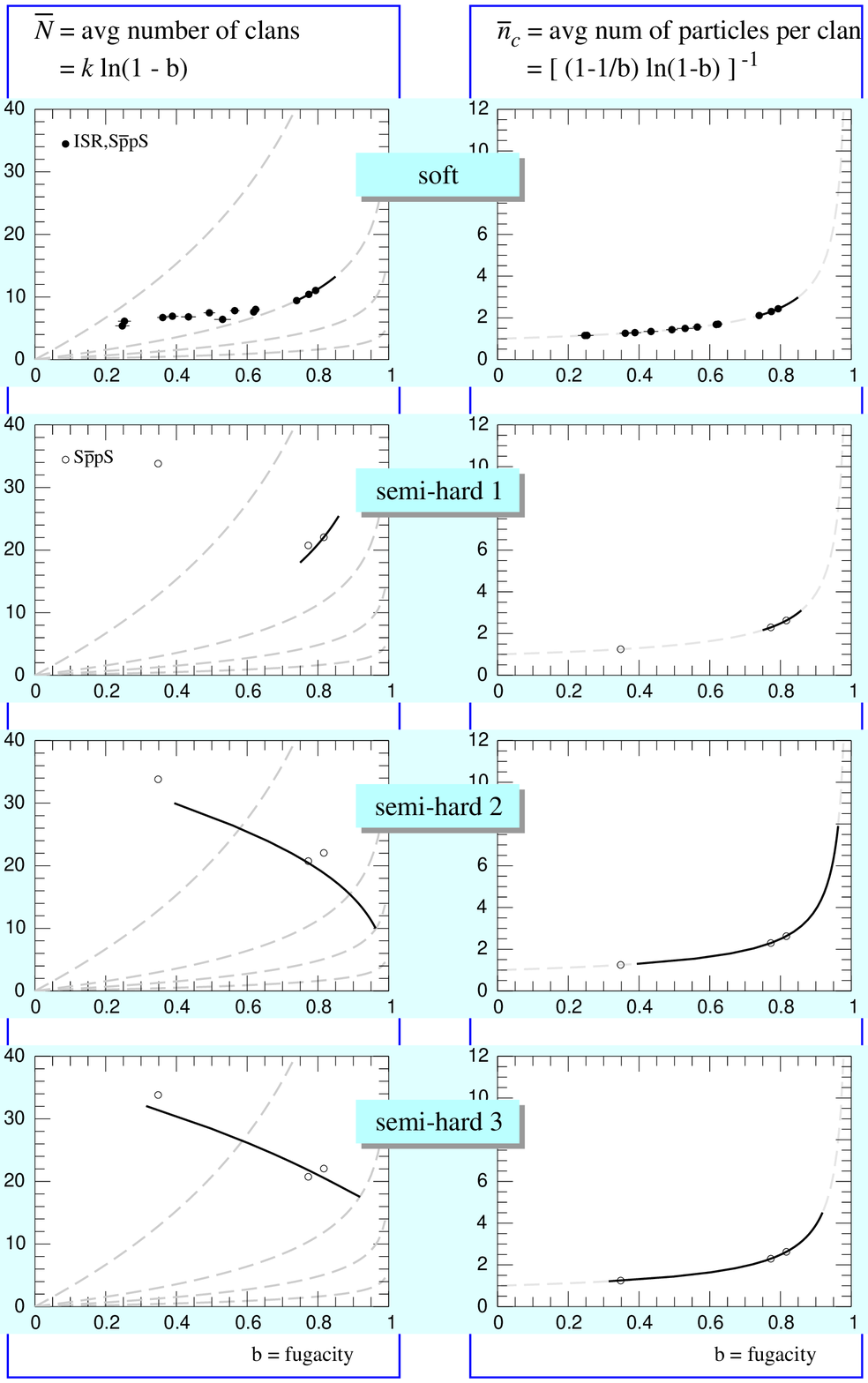}}
  \end{center}
  \caption{Clan parameters dependence on fugacity.}\label{fig:clan}
  \end{figure}

Then we explore the dependence of clan parameters on 
the fugacity $b$, induced by its energy evolution:
\begin{equation}
	\Nbar = k \ln \left( 1 + \frac{\nbar}{k} \right) =
					- k \ln( 1- b);
\end{equation}
\begin{equation}
		\nc = \frac{ \nbar }{ k \ln \left( 1 + \nbar/k \right) } =
					\frac{ b }{ (b-1)\ln(1-b) }.
\end{equation}
Notice that the average number of particles per clan only depends on
the fugacity $b$.
In Figure~\ref{fig:clan} we show for each component and each scenario
the clan parameters as a function of the fugacity. Again, the points
come from fits to experimental data, the solid lines are our extrapolations.
The dashed grey lines show the variation of clan parameters with $b$ at fixed
$k$ (that is, at fixed $V$ and $T$) for the following values of $k$:
1 (lowest curve), 3, 7, 30 (highest curve); being $\nc$ independent of
$k$, only one dashed line is visible in the corresponding graphs.

For the soft and scenario-1 semi-hard components it is
shown that $\Nbar$ is a very slow growing  function of the fugacity of 
the system throughout the ISR region ($b < 0.7$), and
then starts to grow quickly; $\nc$ as a function of the fugacity
has a similar behaviour from $\approx 1.5$ to $\approx 3$.

The decrease of the average number of clans in scenarios 2 and 3 leads
again to the conclusion that this behaviour is closer to that of a
quantum system than to a classical one, favouring as it does the
production of larger clans and therefore of regions of higher particle
density.

For a discussion of other parameters, like the void scaling function,
see again Ref.~\citelow{RU:clanthermo}.

\section{Conclusions}

By defining a new thermodynamical approach to multiparticle production
at parton level we have given the physical meaning of \emph{fugacity}
to a parameter previously used only to describe deviations from
Poisson behaviour in multiplicity distributions.

On this basis, we revisited our previous extrapolations to the TeV
region of inelastic hadron-hadron collisions 
and examine the different behaviours of the two classes of
events (`soft' and `semi-hard').

In the first class, \textit{i.e.}, soft events, the ancestors of the
clans are produced earlier, at higher virtuality and when the
temperature is higher. 
Ancestors in these conditions generate little (clans are small).
This results in a behaviour closer to that of a
classical system (ideal gas).

In the second class, \textit{i.e.}, semi-hard events, the ancestors
are produced later in the cascade, at lower virtualities and when the
temperature is lower.
Ancestors in these conditions are more prolific (clans become larger).
This results in a behaviour closer to that of a quantum system
(stimulated emission); high density regions exist.

Although we used explicitly in the illustration the NB(Pascal)MD,
our result is extensible in principle to any
infinitely divisible distribution which also belongs to
the class of power series distributions.

The results discussed in this paper bring in the spotlight 
the concept of clans, which up to now was only applied in a
statistical framework.  
At this point, it becomes important to investigate
other physical properties of clans, in order to answer questions like
the following ones: can clans be considered
observable objects? if so, what are their quantum numbers?
do they start to interact among themselves in the TeV region? how will
this possibility modify the ideal gas equation of state?

Work in this direction has already begun\cite{RU:clanmass} 
by studying clan masses, with preliminary indications that the 
answer to the first question (observable clans) 
is positive.  This can be extremely relevant for the new heavy
ion machines where the standard examination of events with tens of
thousands of particles may be very problematic.

\section*{References}
\bibliographystyle{prstyR}
\bibliography{abbrevs,bibliography}

\end{document}